\documentclass[lang = american]{ems-icm} 

\usepackage{cleveref}
\newcommand{\dist}{\mathrm{dist}}

\newcommand{\diam}{\mathrm{diam}}
\newcommand{\be}{\begin{equation}}
\newcommand{\ee}{\end{equation}}
\newcommand{\cO}{{\mathcal O}}
\newcommand{\rmd}{{\mathrm d}}
\newcommand{\calD}{{\mathcal D}}
\newcommand{\QL}{Q_{\mathrm{left}}}
\newcommand{\qmax}{q_{\mathrm max}}
\newcommand{\poly}{{\mathrm{poly}}}

\numberwithin{equation}{section}
\newtheorem{lemma}{Lemma}
\newtheorem{theorem}{Theorem}
\begin{document}

\volumetitle{ICM 2022} 

\title{Gapped Quantum Systems: From Higher Dimensional Lieb-Schultz-Mattis to the Quantum Hall Effect}
\titlemark{Gapped Quantum Systems}

\emsauthor{1}{Matthew B. Hastings}{M. B. Hastings}

\emsaffil{1}{Microsoft Quantum and Microsoft Research, Redmond, WA 98052, USA \email{mahastin@microsoft.com}}

\begin{abstract}
We consider many-body quantum systems on a finite lattice, where the Hilbert space is the tensor product of finite-dimensional Hilbert spaces associated with each site, and where the Hamiltonian of the system is a sum of local terms.  We are interested in proving uniform bounds on various properties as the size of the lattice tends to infinity.  An important case is when there is a spectral gap between the lowest state(s) and the rest of the spectrum which persists in this limit,  corresponding to what physicists call a ``phase of matter".  Here,
the combination of elementary Fourier analysis with the technique of Lieb-Robinson bounds (bounds on the velocity of propagation) is surprisingly powerful.  We use this to prove exponential decay of connected correlation functions, a higher-dimensional Lieb-Schultz-Mattis theorem, and a Hall conductance quantization theorem for interacting electrons with disorder.
\end{abstract}

\maketitle

\section{Introduction}
This paper considers lattice quantum systems\footnote{For simplicity, throughout we consider systems where the Hilbert space has a tensor product structure.  It is straightforward to extend these results to the case where fermions obeying canonical anti-commutation relations are present; roughly, this is done by replacing certain commutators with anti-commutators as needed.  We omit this for simplicity in this presentation.}.  It is worth having some concrete examples in mind.  
For one specific example, consider a Hilbert space $({\mathbb C}^{2})^{\otimes L}$, i.e., the tensor product of $L$ Hilbert spaces, each of dimension $2$.  Index these two-dimensional Hilbert spaces (called ``spins" or ``sites") with an integer $i$, taken periodic modulo $L$, and consider
the Hamiltonian
\be
\label{HMG}
H=J_1 \sum_i \vec S_i \cdot \vec S_{i+1}+J_2 \sum_i \vec S_i \cdot \vec S_{i+2},
\ee
where $\vec S_i=(S^x_i,S^y_i,S^z_i)$ denotes the spin operators on the $i$-th such Hilbert space, i.e.,
$$S^x=\begin{pmatrix} 0 & 1/2 \\ 1/2 & 0 \end{pmatrix}, \;
S^y=\begin{pmatrix} 0 & \sqrt{-1}/2 \\ -\sqrt{-1}/2 & 0 \end{pmatrix}, \;
S^z=\begin{pmatrix} 1/2 & 0 \\ 0 & -1/2 \end{pmatrix}.$$
For $J_2=(1/2)J_1>0$, the lowest eigenvalue of $H$ is doubly degenerate for even $L$.   A basis of ground states
consists of either pairing sites $2i,2i+1$ in a singlet for all $i$ or pairing sites $2i,2i-1$ in a singlet; this is called the Majumdar-Ghosh chain\cite{majumdar1969next}.
A slight perturbation of the Hamiltonian, taking $J_2/J_1$ slightly different from $1/2$, breaks the degeneracy of the lowest eigenvalue, but there is an exponentially small difference between the two lowest eigenvalues followed by a gap to the rest of the spectrum that remains nonvanishing as $L\rightarrow \infty$.

This system exemplifies some of the results that we will consider. The fact that for all $J_1,J_2$ the lowest eigenvalue is either degenerate or has vanishing (in the large $L$ limit) difference to the next lowest eigenvalue  is a corollary of the classic Lieb-Schultz-Mattis theorem\cite{lieb1961two}.  That theorem is applicable only to one-dimensional quantum systems with periodic boundaries, meaning that sites can be arranged on a circle with short-range interactions.  We will explain a more general machinery that allows us to prove  a similar theorem for higher-dimensional quantum systems such as those on a two-dimensional square lattice\cite{hastings2004lieb}.  

Also, as $J_1,J_2$ vary, the properties of the Hamiltonian of \cref{HMG} change, but for $J_2$ close to $J_1/2$, the connection correlation functions
are exponentially decaying in the distance between the operators.
Here a connected correlation function is
$\langle A B \rangle - \langle A P B \rangle$ where $A,B$ are operators supported on some set of sites, $P$ projects onto the  two lowest (approximately) degenerate eigenvalues, and $\langle \ldots \rangle$ denotes the expectation value in an eigenvector corresponding to such an eigenvalue.
  This decay follows from another theorem that we will discuss, on exponential decay of correlation functions, again valid in any dimension under some assumptions on the Hamiltonian.

The Hamiltonian of \cref{HMG} obeys a symmetry, namely the Hamiltonian commutes with the three operators $\sum_i S^x_i,\; \sum_i S^y_i$, and  $\sum_i S^z_i$.  Indeed, we will consider often Hamiltonians which just commute with a single operator, such as $\sum_i (S^z_i+1/2)$; here we add a factor $1/2$ so that the eigenvalues of $S^z_i+1/2$ are integers and we can interpret this operator as a ``conserved charge".  Studying such Hamiltonians further in two-dimensions leads to the question of quantum Hall conductance, also discussed here.

Surprisingly, the key to many of these results is to consider the dynamical properties of the system, where we consider correlation functions of operators at different times, using the technique of Lieb-Robinson bounds.
Let us begin by defining the systems we consider in some generality.

\subsection{Lattice Quantum Systems}
We consider quantum systems defined on a finite lattice.  We have some finite set $\Lambda$ of \emph{sites}.  The set $\Lambda$ is called the lattice. 
Associated with each site $i$ is some finite-dimensional Hilbert space, and the Hilbert space of the whole quantum system is the tensor product of these Hilbert spaces.
There is some metric $\dist(i,j)$ where $i,j\in \Lambda$.
In many applications $\Lambda$ may indeed be a geometric lattice in $\mathbb{R}^n$ or in $\mathbb{T}^n$ for some $n$; in this case, we call the system $n$-dimensional, and in this case the metric is inherited from the ambient space $\mathbb{R}^n$ or $\mathbb{T}^n$.  However, in general $\Lambda$ may be an arbitrary set with arbitrary metric.

The Hamiltonian $H$ has the form
\be
\label{Hdef}
H=\sum_X h_X,
\ee
where the sum ranges over $X\subset \Lambda$, and each $h_X$ is self-adjoint and is supported on set $X\subset \Lambda$.

We use $\Vert O \Vert$ to denote the operator norm (largest singular value) of $O$.
Our interest is in local Hamiltonians; locality
is expressed as some assumption on the norms $\Vert h_X\Vert$ as a function of the diameter of the sets $X$. 

One typical assumption is that the Hamiltonian has \emph{bounded strength and range} and that the set of sites $\Lambda$ has \emph{bounded local geometry}, meaning that all $\Vert h_X \Vert \leq J$ for some \emph{strength} $J$ and all sets $X$ has $\diam(X)\leq R$ for some \emph{range} $R$ and that for all $i\in \Lambda$, we have $|\{j\in \Lambda|\dist(i,j)\leq R\}|$ bounded by some constant.  Other assumptions considered include exponential decay where $\Vert h_X \Vert$ is exponentially small in $\diam(X)$.

From certain such locality assumptions, one can derive a so-called
Lieb-Robinson bound, which can be thought of as bounding the velocity of excitations in
such a lattice quantum system.
For an arbitrary operator $A$, let
\be
A(t) \equiv \exp(i H t) A \exp(-i H t)
\ee
denote the operator $A$ evolved for time $t$ under Hamiltonian $H$.

The first such bound was proven by Lieb and Robinson\cite{lieb1972finite}.  However, their proof gave bounds that depended on the dimension of
the Hilbert space.  In \cref{LRapp} we give a different proof that does not depend on the dimension, following the strategy in \cite{hastings2004lieb} as slightly modified in \cite{hastings2006spectral}.  Indeed, rather than proving a specific bound,
we give a series expansion (\ref{CBseries}) below which upper bounds
$\Vert[A(t),B]\Vert$ and different assumptions on $\Vert h_X\Vert$ can be inserted into this series. 
Using this series expansion,
a typical result\cite{hastings2006spectral} is
\begin{lemma}
\label{lrlem}
Suppose there are positive constants $\mu,s$ such that for all sites $i$ we have
$$\sum_{X\ni i} \Vert h_X \Vert |X| \exp[\mu \diam(X)]\leq s <\infty.$$
Then,
for any sets $X,Y$ with $\dist(X,Y)>0$, and any operators $A,B$ supported on $X,Y$ respectively, 
\begin{eqnarray*}\Vert [A(t),B]\Vert&\leq&
 2\Vert A\Vert \Vert B\Vert \sum_{i \in X} \exp[-\mu\dist(i,Y)]\left[e^{2s|t|}-1\right].\end{eqnarray*}

As a corollary,
defining $v_{LR}=4s/\mu$, for $\dist(X,Y)\geq v_{LR} t$, we have
$\Vert [A(t),B]\Vert \leq |X| \cdot \Vert A \Vert \Vert B \Vert \exp[-\frac{\mu}{2}\dist(i,Y)]$.
\end{lemma}
This quantity $v_{LR}$, called the Lieb-Robinson velocity, can be thought of as defining a ``light-cone"\cite{bravyi2006lieb}, so that up to exponentially small error $A(t)$ is supported within distance $v_{LR} t$ of $X$.

Note that \cref{lrlem} is applicable to the case of bounded strength and range with bounded local geometry.

Remark: the fact that the commutator is not vanishing, but merely very small, outside the light-cone is sometimes called ``leakage".
In most applications of these bounds, the leakage is negligibly small compared to other terms.  Indeed, for the rest of the this paper, the leakage terms will be negligible and we will avoid any detailed discussion of them.  However, we emphasize that the leakage really is non-zero in every case of interest; since the commutator is an analytic function of time (this follows trivially since we have a finite-sized system), if the commutator were exactly zero on some interval of time, then it would vanish for all times.

We emphasize that we consider finite size systems, so that many properties can be defined in an elementary way.  For example, the Hamiltonian is a finite-dimensional matrix; the ground state energy is simply the smallest eigenvalue of the Hamiltonian; if the smallest eigenvalue is non-degenerate, then the ground state is simply the corresponding eigenvector (up to some arbitrary phase); and correlation functions are simply the trace of the projector onto the ground state with some given other finite-dimensional matrices.  This contrasts with
considering systems directly in the infinite size limit where one must take some care to define an algebra of operators on an infinite system.
However, although we consider finite-size systems, our interest is in bounds that are uniform in $|\Lambda|$.

\subsection{Outline of Results and Notation}
We will survey some of the results that have been obtained using these methods.  A key role is played by the \emph{spectral gap}.  In this paper, unless stated otherwise, the spectral gap is defined to be the absolute value of the difference between the ground state energy of $H$ (assumed non-degenerate) and the next smallest eigenvalue.  We denote the spectral gap $\Delta E$.

We will be loose about estimates.  In many cases we will simply state that a term is small (perhaps exponentially small or some other decay), leaving the detailed proofs and precise bounds to the already-published literature.  This is done to emphasize the ideas without getting too involved in the estimates.
At the same time, we will give examples from physics to motivate the constructions.

We use computer-science big-O notation such as $\Omega,\cO,\ldots$ throughout, where we implicitly consider a family of Hamiltonians defined on a family of $\Lambda$ with increasing cardinality $\Lambda$.  The control parameter for the big-O notation may be $|\Lambda|$ or in the higher dimensional Lieb-Schultz-Mattis theorem later may be some other distance scale.  

We use $|\ldots|$ for $\ell_2$ norm of a state vector.  We use $I$ for the identity matrix.  We use $A^\dagger$ to denote the Hermitian conjugate of an operator $A$.

When we refer to a quantum state, this will always be a normalized pure state (with an arbitrary phase), rather than a mixed state.

In \cref{cluster}, we sketch the proof that connected correlation functions decay exponentially in systems with a spectral gap\cite{hastings2004lieb,hastings2006spectral}.
This can be understood as a non-relativistic analogue of a familiar result in relativistic quantum field theory (where the speed of light plays the role of $v_{LR}$) that correlation functions decay exponentially in gapped theories.
In \cref{lsm}, we sketch the proof of the higher dimensional Lieb-Schultz-Matthis theorem, proven in \cite{hastings2004lieb}.  In \cref{qhe}, we sketch the proof of Hall conductance quantization\cite{hastings2015quantization}.  These last two results have a certain topological flavor.  In both cases, one of the physical ideas motivating the mathematical proof is that although correlation functions decay exponentially in gapped systems, it is still possible for there to be some kind of topological order in the ground state.

\section{Exponential Decay of Connected Correlation Functions}
\label{cluster}
In massive relativistic quantum field theories, connected correlation functions decay exponentially.  Here we consider similar results for lattice field theories.  The Lieb-Robinson velocity plays some role similar to that of the speed of light in a relativistic field theory, while the spectral gap plays a role similar to that of a mass gap.

A typical result for exponential decay is
\begin{theorem}
Let $A_X,B_Y$ be supported on sets $X,Y$.  Suppose the conditions of \cref{lrlem} hold.  Assume there is a unique ground state with spectral
 gap $\Delta E$ to the rest of the spectrum.  Let $\langle \ldots \rangle$ denote the expectation value in the ground state.  Then,
$$\langle A_X B_Y \rangle - \langle A_X \rangle \langle B_Y \rangle
\leq
\Vert A_X \Vert \cdot \Vert B_Y \Vert
\Bigl(\exp\Bigl[-\Omega\Bigl(\frac{\dist(X,Y) \Delta E}{v_{LR}}\Bigr)\Bigr]+\ldots\Bigr),$$
where $\ldots$ denotes a leakage
term from the Lieb-Robinson bound which is bounded by $|X|\cdot \Vert A_X \Vert \Vert B_Y \Vert$ times an exponentially
decaying function of $\dist(X,Y)$.
\end{theorem}

Remark: This result can be readily generalized to the case that
$H$ has a $q$-fold degenerate (or almost degenerate) smallest eigenvalue and then a gap $\Delta E$ to the rest of the spectrum.  Then, defining $P_0$ to project onto the ground state subspace and $\langle O \rangle\equiv \frac{1}{q}{\mathrm tr}(P_0 O)$, one may derive a more general bound on
$\langle A_X B_Y\rangle - \langle A_X P_0 B_Y \rangle$.  In this case, there is an additional term in the bound that vanishes in the limit that the $q$ lowest eigenvalues become exactly degenerate.

Remark: the proof follows a general outline that we will use to derive the later results also, and after giving the proof we will emphasize the ideas that will be repeated later.

{\emph Proof of theorem:} To ease notation, replace $A_X$ by $A_X - \langle A_X\rangle$ and $B_Y$ by $B_Y-\langle B_Y \rangle$, so that both $A_X,B_Y$ have vanishing expectation value in the ground state.  
Also, let $\ell=\dist(X,Y)$

Let $\Psi_n$ for $n=0,1,\ldots$ be an orthonormal basis of eigenstates of $H$, with eigenvalues $E_0<E_1\leq \ldots$.  Let $\langle \psi,\phi\rangle$ denote the inner product between state $\psi,\phi$.
Then, 
\begin{align}
\label{nfeq}
\langle [A_X(t) ,B_Y]\rangle=&\sum_{n>0} \langle \psi_0,A_X \Psi_n\rangle \langle \Psi_n, B_Y \psi_0\rangle \exp(-i (E_n-E_0) t)
\\ \nonumber
&-\sum_{n>0} \langle \psi_0, B_Y \psi_n\rangle \langle \psi_n, A_X \psi_0\rangle \exp(+i (E_n-E_0) t).
\end{align}
Thus, to compute the desired correlation function $\langle A_X(t=0) B_Y\rangle$, we want to extract the ``negative frequency" part of $\langle [A_X(t) ,B_Y]\rangle$, meaning the first sum in \cref{nfeq}, evaluated at $t=0$.  To do this, we use the following lemma\cite{hastings2004lieb,hastings2006spectral}.  This lemma shows a typical kind of technique in this subject: we have some function (in this case a step function) which has a singularity, and we construct some other function (or in this case, a limit of a family of functions) which is a good approximation to that function when the argument has absolute value $\geq \Delta E$, and we show that that approximation has a fast decaying Fourier transform.

\begin{lemma}
\label{lemma:Hastings}
Let $E\in{\mathbb R}$, and $\alpha>0$. Then 
\begin{align}\nonumber
\lim_{T\uparrow\infty}\lim_{\epsilon\downarrow 0}
\frac{i}{2\pi}\int_{-T}^T\frac{e^{-iEt}e^{-\alpha t^2}}{t+i\epsilon}\rmd t
&=\frac{1}{2\pi}\sqrt{\frac{\pi}{\alpha}}
\int_{-\infty}^0\rmd \omega\exp[-(\omega+E)^2/(4\alpha)]\\
&=\begin{cases} 1+\cO(\exp[-\Delta E^2/(4\alpha)]) \; & {\text for}  \; E\ge \Delta E;\\ \nonumber
\cO(\exp[-\Delta E^2/(4\alpha)])\; &{\text for} \; E\le -\Delta E.
\end{cases}
\end{align}
\end{lemma}

Using \cref{lemma:Hastings} one has 
\begin{align}
\label{extractTcorr} 
&\left\langle\Phi,A_X B_Y\Phi\right\rangle
\\ \nonumber  =&\lim_{T\uparrow\infty}\lim_{\epsilon\downarrow 0}\frac{i}{2\pi}\int_{-T}^T \rmd t 
\frac{1}{t+i\epsilon}
\left\langle[A_X(t),B_Y]\right\rangle e^{-\alpha t^2}+
\cO(\exp[-\Delta E^2/(4\alpha)]).
\end{align}

Now we choose $\alpha$ and apply the Lieb-Robinson bound.  Fix $\alpha=\Delta E v_{LR}/(2 \ell)$.  Then, $\cO(\exp[-\Delta E^2/(4\alpha)])=\cO(\exp[-\ell \Delta E/(2 v_{LR})])$.
This bounds the second term on the right-hand side of \cref{extractTcorr}.
To bound the first 
term, we break the integral over $t$ into an integral for $|t|\leq \ell/v_{LR}$ 
and an integral for $|t|\geq \ell/v_{LR}$.  The integral for $|t|\leq \ell/v_{LR}$
can be bounded using the Lieb-Robinson bound, giving the leakage term in the theorem.  The integral for $|t|\geq \ell/v_{LR}$ is bounded by a triangle inequality by
$$2 \Vert A_X \Vert \cdot \Vert B_Y \Vert \lim_{T\uparrow\infty}\lim_{\epsilon\downarrow 0}\frac{1}{2\pi}\int_{\ell/v_{LR}\leq |t| \leq T} \rmd t 
\frac{1}{t+i\epsilon}
e^{-\alpha t^2},$$
which is bounded by $2 \Vert A_X \Vert \cdot \Vert B_Y \Vert \cdot \cO(\exp[-\ell \Delta E/(2 v_{LR})])$.
\qed

\section{Higher Dimensional Lieb-Schultz-Mattis}
\label{lsm}
\subsection{Review of One-Dimensional Lieb-Schultz-Mattis Theorem}
One-dimensional quantum spin systems with $SU(2)$-invariant Hamiltonians exhibit different behavior depending on whether the spin is integer or half-integer.  For half-integer spin, it is found that either the ground state is degenerate or the gap vanishes in the thermodynamic limit, while for integer spin there may be a unique ground state with a gap.

One paradigmatic example is the spin-$S$ Heisenberg spin chain,
$$H=J \sum_i \vec S_i \cdot \vec S_{i+1},$$
$J>0$,
where here the sites are labelled by integers $i=0,1,\ldots,L-1$, which are periodic modulo $L$, and corresponding to each site there is a $2S+1$-dimensional Hilbert space corresponding to a spin-$S$ representation of $SU(2)$, and where $\vec S_i$ is a vector of spin operators on the $i$-th spin.
For $J>0$ and spin-$1/2$,
there is a continuous spectrum of excitations.  The spectral gap then vanishes polynomially in $L$.
Another paradigmatic example is the Majumdar-Ghosh chain mentioned at the start of this paper.
On the other hand, for integer spin the famous ``Haldane conjecture"\cite{haldane1983continuum} argues that there is a  unique ground state with a spectral gap which is $\Omega(1)$, and the AKLT Hamiltonian\cite{affleck1987rigorous} is an exactly solvable spin-$1$ Hamiltonian which shows this property.

This vanishing of the gap for half-integer spins is a corollary of 
the Lieb-Schultz-Mattis theorem\cite{lieb1961two,affleck1986proof} in one-dimension.  We will give the theorem in
a more general setting where the Hamiltonian is $U(1)$-symmetric, without using the full $SU(2)$ symmetry, and then relate this to the case of spin chains.

Let us define some terms.  Say that a system is one-dimensional with periodic boundary conditions of size $L$, for some integer $L$, if the sites in $\Lambda$ correspond to vertices of a cycle graph with $|\Lambda|=L$, with the shortest path metric on the graph being the distance, and if all sites have the same Hilbert space dimension.  We will label sites by integers.
Define a translation operator $T$ in the obvious way; $T$ is a unitary operator and conjugation by $T$ maps the algebra of operators supported on site $i$ to those supported on site $i+1$, and $T^L=I$.
We say that a Hamiltonian is translationally invariant if $T H T^{-1}=T$.
We say that a system has a conserved charge $Q$ if $Q=\sum_i q_i$, with $q_i$ is an operator with integer eigenvalues supported on site $i$ with $\Vert q_i \Vert \leq \qmax$ for some $\qmax$, so that $q_j=T^{j-i} q_i T^{i-j}$, so that
$$[Q,H]=0.$$

The proof that follows uses a trick of averaging over two choices of twist, to use the minimum number of assumptions; this form of the proof seems to first appear in \cite{koma2000spectral}.
\begin{theorem}
Consider a one-dimensional system with periodic boundary conditions of size $L$ with translation invariant Hamiltonian $H$ and a conserved charge $Q$.
Further, assume $H$ has strength $J$ and range $R$.

Let $\psi_0$ be a ground state of $H$ with $|\psi_0|=1$, and suppose that
$\frac{\langle \psi_0,Q \psi_0\rangle}{L}$ is not integer.
Then, if the ground state is non-degenerate, the spectral gap of $H$ is bounded by
$\cO(\frac{J\qmax^2R^2}{L})$.
\begin{proof}
The proof is variational: construct another state,  show that the state is orthogonal to $\psi_0$, and compute the expectation value of $H$ in this state.
Let
$$U_{LSM}=\exp(\pm i A),$$
where we will pick the sign later and
where
$$A \equiv \sum_{j=0}^{L-1} 2\pi q_j \frac{j}{L}.$$
The variational state used is
$$\Phi=U_{LSM} \psi_0.$$

Since the ground state of $H$ is non-degenerate and  since $[T,H]=0$, we have $T\psi_0=z\psi_0$ for some scalar $z$ with $|z|=1$.
Note that
\begin{align}
T\Phi&=T\exp[\pm i \sum_{j=0}^{L-1} 2\pi q_j \frac{j}{L}] \psi_0\\ \nonumber
&=\exp[\pm i \sum_{j=0}^{L-1} 2\pi \Bigl(T q_j T^{-1}\Bigr) \frac{j}{L}] T\psi_0 \\ \nonumber
&= z\exp[\pm i \sum_{j=0}^{L-1} 2\pi \Bigl(T q_j T^{-1}\Bigr) \frac{j}{L}] \psi_0\\ \nonumber
&= z\exp[\pm i \sum_{j=0}^{L-1} 2\pi q_{j+1} \frac{j}{L}]  \psi_0 \\ \nonumber
&= z\exp[\pm i \sum_{j=0}^{L-1} 2\pi q_{j} \frac{j-1}{L}]  \psi_0 \\ \nonumber
&= z \exp[\mp i 2\pi \frac{Q}{L}] \Phi \\ \nonumber
&=z \exp[\mp i 2\pi \frac{\langle \psi_0,Q \psi_0\rangle}{L}] \Phi.
\end{align}
The equality on the fifth line is a change in the index of summation, replacing $j$ by $j-1$.  Here the assumption that $q_j$ has integer eigenvalues is used so that $\exp[i 2\pi q_L \frac{L}{L}]=\exp[i 2\pi q_0 \frac{0}{L}]=I$.
The equality on the final line uses the assumption that $[Q,H]=0$ so that $\psi_0$ is an eigenvector of $Q$.

Using the assumption that $\langle \psi_0,Q \psi_0\rangle/L$ is non-integer, it follows that $\Phi$ is an eigenvector of $T$ with eigenvalue different from $z$ so it is orthogonal to $\psi_0$.

Now we estimate the energy of this state.
Write $H=\sum_{i=0}^{L-1} h_i$, with $h_j=T^{j-i} h_i T^{i-j}$ and with each $h_i$ supported on the set of sites within distance $R$ of $i$, with $\Vert h_i \Vert \leq J$.

We average $\langle \Phi, H \Phi\rangle-\langle \psi,H \psi \rangle$ over the two choices of sign in $U_{LSM}$, giving
\begin{align}\nonumber
&\langle \Psi,\Bigl( \frac{U_{LSM}^\dagger H U_{LSM}+U_{LSM}H U_{LSM}^\dagger}{2}- H\Bigr) \Phi\rangle 
\\ \nonumber
=&
L \langle \Psi,\Bigl( \frac{U_{LSM}^\dagger h_0 U_{LSM}+U_{LSM}h_0 U_{LSM}^\dagger}{2}- h_0\Bigr) \Phi\rangle
\\ \nonumber
\leq &  L \Bigl\Vert \frac{U_{LSM}^\dagger h_0 U_{LSM}+U_{LSM}h_0 U_{LSM}^\dagger}{2}- h_0 \Bigr\Vert \\ \nonumber
\leq &  L \Vert [[h_0,A],A] \Vert,
\end{align}
where the averaging over signs cancels terms $[h_0,A]$.
Finally, $\Vert [[h_0,A],A] \Vert=\cO(\frac{J\qmax^2R^2}{L^2})$.
\end{proof}
\end{theorem}

To apply this system to $SU(2)$-invariant spin chains with half-integer spin, we may take $q_i=1/2+S^z_i$, where $S^z_i$ is the $z$-component of the $i$-th spin.  Then, if the ground state is non-degenerate it has total spin $0$, and hence $\langle \sum_i S^z_i\rangle=0$, so $\langle Q \rangle/L=1/2$, non-integer.

It is instructive to consider this variational state in the case of the Majumdar-Ghosh chain \cref{HMG}.  A basis of ground states corresponds to pairing neighboring sites in singlets in one of two ways.
Taking the sum of these states gives an eigenvector of $T$ with eigenvalue $+1$.  Applying $U_{LSM}$ to this sum gives the difference of these two states, up to an error of order $\cO(1/L)$.  The difference of these states is an eigenvector $T$ with eigenvalue $-1$.

\subsection{Higher Dimensional Extensions: Physics}
One might try to extend this theorem beyond one-dimensional systems.
Note first that translation invariance is necessary for the theorem to hold: one can easily construct spin-$1/2$ systems without translation invariance with a unique ground state and a gap such as
$$H=\sum_i \vec S_{2i} \cdot \vec S_{2i+1}.$$

The higher dimensional theorem will apply to $n$-dimensional quantum systems, for $n\geq 1$.
However, we will be able to state the theorem in greater generality, which will also be convenient because it will emphasize the fact that we use translation invariance in only one direction.

Say that a system has \emph{translation invariance in one direction with periodicity} $L$ if the sites can be
labelled by a pair $(i,v)$, where $i$ is an integer labelling a vertex of a cycle graph of length $L$ and $v$ is a vertex in some other graph $G$, so that the following hold.
First, the metric is the shortest path metric on the graph given by the Cartesian product of that cycle graph with $G$.
Second, the Hilbert space dimension of site $(i,v)$ is some $d_v$ depending only on $v$.
Given this, define
a unitary operator $T$ such that conjugation by $T$ maps the algebra of operators supported on site $(i,v)$ to those supported on site $(i+1,v)$ for all $v$, and $T^L=I$.
We say that a Hamiltonian is translationally invariant if $T H T^{-1}=T$.
We say that a system has a conserved charge $Q$ if $Q=\sum_{i,v} q_{i,v}$, with $q_{i,v}$ is an operator with integer eigenvalues supported on site $i$
with $\Vert q_{i,v} \Vert \leq \qmax$, so that $q_{j,v}=T^{j-i} q_{i,v} T^{i-j}$, so that
$$[Q,H]=0$$.

For an $n$-dimensional quantum system,
$G$ may be a $(n-1)$-fold Cartesian product of cycle graphs, so that sites are labelled by $n$ different integers, each periodic modulo some other integer.

Then it follows that\cite{hastings2004lieb}
\begin{theorem}
\label{hlsm}
Consider a system with
translation invariance in one direction with periodicity $L$,
with a Hamiltonian with strength and range $J,R$ both $\cO(1)$, and such that $\Lambda$ has bounded local geometry.  Assume that the number of sites is $\cO(\poly(L))$.  Assume that there is a conserved charge with $\qmax=\cO(1)$.
Assume that the ground state $\psi_0$ is unique with
$\langle \psi_0, Q \psi_0\rangle/L$ non-integer.
Then, the gap $\Delta E$ is
$\cO(\log(L)/L)$,
where the constants hidden in the big-O notation depend on $J,R,\qmax$, on the polynomial bounding the number of sites, and on the local geometry of $G$.
\end{theorem}

Note that the theorem is slightly weaker than in the one-dimensional case, with a bound $\cO(\log(L)/L)$ rather than $\cO(1/L)$. Also, for simplicity, we have been less explicit about the dependence of the bound on the constants.

Note also one slightly unsatisfactory feature: the theorem requires that $\langle \psi_0, Q \psi_0\rangle/L$ be non-integer.  Suppose that we consider a  two-dimensional system, of size $L$-by-$L'$, with $\langle \psi_0, Q \psi_0\rangle/(L L')=1/2$.  This is a typical case of interest in spin systems.  Then, the theorem is only applicable if $L'$ is odd.

One might attempt to use the same proof as before.  Suppose that $H$ has bounded strength and range.
Applying the same variational argument, the change in the expectation value of every term in the Hamiltonian (e.g., $\langle \Phi,h_X\Phi\rangle - \langle \psi_0, h_X \psi_0\rangle$)
is still $\cO(1/L^2)$, but the number of such terms now is not $L$ but rather proportional to $L$ times the number of vertices in $G$.  As a typical application of interest, let $L_x=L$ and let $G$ be a cycle graph of length $L_y$ so that
vertices are labelled by a pair of integers $(i,j)$, with $i$ periodic modulo $L_x$ and $j$ periodic modulo $L_y$.  Then if the ``aspect ratio" $L_y/L_x$ is of order unity,
the variational state may have energy of order unity above the ground state\cite{affleck1988spin}.

The problems with this approach were given in a very insightful physics article\cite{misguich2000some}.
Indeed, the problem is not purely mathematical.  The problem is that a two dimensional quantum spin system can exhibit completely different behavior from a one-dimensional quantum spin system.  A one-dimensional system with spin-$1/2$ can have a state like the Heisenberg chain with a polynomially small gap and power-law decaying correlations.  Alternatively, it can have a state like the Majumdar-Ghosh chain.  In this exactly solvable example, two choices of ground state correspond to two different ways of pairing neighbors into singlets, either pairing site $2i$ with $2i+1$ or pairing site $2i$ with $2i-1$.  These two choices each break translation symmetry, though one may take symmetric and anti-symmetric combinations to obtain ground states which are eigenstates of $T$.  There is a local order parameter which distinguishes between these states\footnote{See \cite{affleck1986proof} for results showing that, in a sense, these are the only two possibilities for a one-dimensional system, either translational symmetry breaking or a continuous spectrum.}, i.e., indeed, there is an operator supported on a set of bounded diameter
which has non-vanishing matrix elements between symmetric and anti-symmetric states.
However, in two dimensions, there is a new possibility.  Spins can pair into singlets (or dimers as they are called) but these dimers can enter into a quantum spin liquid state\cite{sutherland1988systems,rokhsar1988superconductivity,read1989statistics}. In this case, the physics is that there is still a ``topological degeneracy", so that there is exponentially small splitting between the two lowest eigenstates.
However, there is no local order parameter: any operator supported on a set of bounded diameter is exponentially close to a scalar in the subspace of the two lowest eigenstates.

So, while these physics arguments provide some motivation to find a generalization of the Lieb-Schultz-Mattis theorem to higher dimensional systems, they also show that the variational argument does not directly generalize.  To prove the theorem, we need the tool of ``quasi-adiabatic continuation", described in the next subsection.

This tool is used to turn physics arguments based on an idea of twisting boundary conditions into precise results.

Assume $H$ has finite range $R$, with $L>>R$.

\begin{figure}
\includegraphics[width=2in]{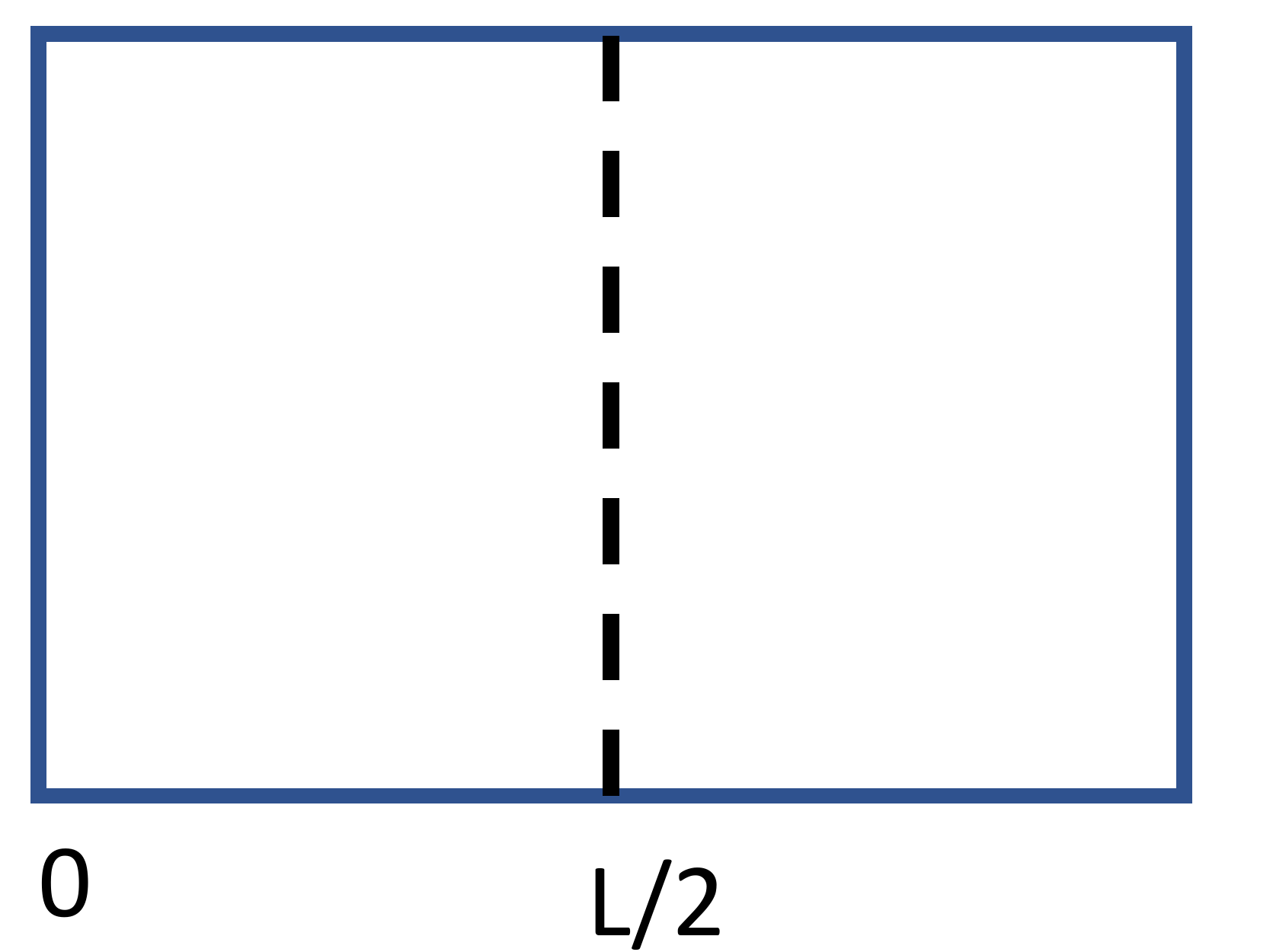}
\caption{Schematic illustration of two-dimensional system.  Individual sites are not shown.  Left and right sides of the system are identified to give a cylinder (or torus if top and bottom are also identified).  Numbers indicate first coordinate.  The operator $\QL$ is the total charge on sites between the left side and the dashed line.  Twist $\theta$ twists terms near the left side of the figure, while $\theta'$ twists terms near the dashed line.  The twist $\theta'$ is used as a technical trick in the proof of the higher-dimensional Lieb-Schultz-Mattis theorem.}
\label{figlsm}
\end{figure}

Given any site $(i,v)$, its \emph{first coordinate} is the integer $i$.
Let $$\QL=\sum_{0\leq i < L/2} \sum_v q_{i,v}$$
be the total charge on sites with first coordinate $0\leq i < L/2$.  This is the ``left" half of the system in \cref{figlsm}.

Say that a set has first coordinate \emph{near} $i$ if it is within distance $R$ of the set of sites with first coordinate $i$, treating the first coordinate periodic modulo $L$.
Define a two-parameter family of Hamiltonians (note the signs in the exponents)
\begin{align}\nonumber H_{\theta,\theta'}\equiv& \sum_{X \, \mathrm{near} \, 0} \exp(i \theta \QL) h_X \exp(-i \theta \QL)+
\sum_{X \, \mathrm{near} \, L/2} \exp(-i \theta' \QL) h_X \exp(i \theta' \QL)
\\\nonumber &+\sum_{\mathrm{remaining}\; X} h_X,\end{align}
where $\theta,\theta'$ are real parameters, periodic modulo $2\pi$.
The last sum is over $X$ not near $0$ or $L/2$.
Note that
$$H=H_{0,0}$$
and
\be
\label{gaugeinvar}
H_{\theta+\phi,\theta'-\phi}=\exp(i \QL \phi) H_{\theta,\theta'} \exp(-i \QL \phi).
\ee

Physically, we can regard $\theta$ as defining a ``gauge field" along the line of sites with first coordinate $0$, and $\theta'$ as another gauge field along the line $L/2$.  Then
\cref{gaugeinvar} describes a gauge transformation relating two different ``choices of gauge", with the sum $\theta+\theta'$ invariant under this.  While this relation may seem trivial, it will be very useful in what follows.

Oshikawa\cite{oshikawa2000commensurability} considered the effect of changing $\theta$ from $0$ to $2\pi$ in an attempt to prove a higher-dimensional Lieb-Schultz-Mattis theorem.  
In \cite{misguich2000some}, his argument was analyzed in more detail, where it was found that what it proves is that the ground state must become degenerate at some value of $\theta$
if $\langle \psi_0,Q \psi_0\rangle/L$ is non-integer.  Proof: suppose the ground state is non-degenerate for all $\theta$.  Then, adiabatic evolution from $\theta=0$ to $2\pi$ maps the ground state to itself.  However, one may show that if the ground state is an eigenvector of the translation operator $T$ with eigenvalue $z$, the adiabatically evolved state has eigenvalue $\exp(i \frac{2\pi}{L}\langle \psi_0,Q \psi_0\rangle) z\neq z$.

This kind of argument considering a change in boundary conditions is called flux insertion, and is related to Laughlin's argument for Hall conductance quantization\cite{laughlin1981quantized}.  While adiabatic evolution using flux insertion does not prove the desired result, this is still a useful physical idea.
The technical tool we use to make the idea of flux insertion rigorous is called ``quasi-adiabatic continuation"\cite{hastings2004lieb}.

\subsection{Quasi-Adiabatic Continuation}
Suppose we have some family of Hamiltonians $H_s$ which depend smoothly on a real parameter $s$.  Assume that for all $s$ the ground state is non-degenerate and the spectral gap is $\geq \Delta E$.  Let $\psi_a(s)$ for $a\geq 0$ denote (orthonormal) eigenstates of $H_s$, with energies $E_a(s)$, with $\psi_0(s)$ being the ground state.

Then, a familiar result of first order perturbation theory is that
\be
\partial_s \psi_0(s)=\sum_{a>0}   \frac{1}{E_0(s)-E_a(s)}  \psi_a(s) \langle \psi_a(s), \Bigl( \partial_s H_s \Bigr)  \psi_0(s)\rangle.
\ee
We now use a trick similar to that used in the proof of exponential decay of correlations above: we take some function of energy difference which has some singularity, i.e., in this case, $1/(E_0-E_a)$, and we approximate that function by a smooth function which gives a good approximation when the energy difference is $\geq \Delta E$.  In the case of exponential decay of correlations, the needed result is \cref{lemma:Hastings}.
Here a variety of different forms have been used, and rather than giving details we simply give the approach in outline.

Let $f(\cdot)$ be some smooth function with some Fourier transform $\tilde f(\cdot)$.  Assume $f(x) \approx -1/x$ for $|x|\geq \Delta$ and $f(0)=0$ and $f(x)=-f(-x)$, where we will not be precise about the meaning of the approximation $\approx$.  
Then,
\begin{align}
\label{qaddefeq}
\partial_s \psi_0(s)& \approx \sum_{a>0}   f(E_a-E_0)  \psi_a(s) \langle
 \psi_a(s) ,\Bigl( \partial_s H_s \Bigr)  \psi_0(s)\rangle \\ \nonumber
&=\sum_a \int_{-\infty}^{\infty} \frac{\rmd t}{2\pi} \, \tilde f(t) \exp(i (E_a-E_0) t)\psi_a(s)  \langle \psi_a(s),
\Bigl( \partial_s H_s \Bigr)  \psi_0(s)\rangle 
\\ \nonumber
&=\int_{-\infty}^{\infty} \frac{\rmd t}{2\pi} \, \tilde f(t)  \Bigl( \exp(+i H_s(t)) \Bigl(\partial_s H_s\Bigr) \exp(-i H_s(t))\Bigr)  \psi_0(s) \\ \nonumber
&\equiv  i \calD_s \psi_0(s),
\end{align}
where the last line of the equation is interpreted as defining an operator $i\calD_s$ called the quasi-adiabatic continuation operator.
The error in this approximation in \cref{qaddefeq} depends on the error in the approximation $f(x) \approx -1/x$ and depends on the norm $\Vert \partial_s H_s\Vert$ and we do not go into details here.

Since we have chosen $f$ to be odd, $\calD_s$ is Hermitian.
We can integrate this quasi-adiabatic continuation operator along a path such as $s\in [0,1]$ to give
$$\psi_1(s) \approx {\mathcal P} \exp\Bigl(i \int_0^1 \calD_s \rmd s\Bigr) \psi_0(s),$$
where $\mathcal P$ denotes a path-ordered exponential and
${\mathcal P} \exp(i \int_0^1 \calD_s \rmd s)$ is a unitary.

The essential point of \cref{qaddefeq} is that if we choose $f$ so that $\tilde f$ is sufficiently rapidly decaying in time, then (by the Lieb-Robinson bounds) the operator $\calD_s$ enjoys certain locality properties.  In particular, if $\partial_s H_s$ is supported on some given set (for example, $\partial_\theta H_{\theta,0}$ is supported within $O(1)$ of the line $0$), then $\calD_s$ can be approximated by an operator supported within some distance $\ell$ that set, with the error in approximation depending on the choice of $\tilde f$, and decreasing as $\ell$ is increased.  Further, if $\partial_s H_s$ is a sum of operators supported on given sets then, by linearity, $\calD_s$ can be approximated by a sum of operators supported within some distances of those sets.

In the original application of quasi-adiabatic continuation\cite{hastings2004lieb}, there were two sources of error.  One source of error came from the approximation in \cref{qaddefeq} since $f(x)$ was not exactly equal to $-1/x$ for $x\geq \Delta$, while the second source of error came from the approximate locality 
of $\calD_s$. 

In \cite{osborne2007simulating}, Osborne introduced a different ``exact" version where $f(x)$ was exactly equal to $-1/x$ for $|x|\geq \Delta E$, and he showed that one could choose $\tilde f$ to decay superpolynomially in time.  Using an old result in analysis\cite{ingham1934note}, it is possible to improve this superpolynomial decay to an ``almost exponential decay", made more precise later.

The original formulation of quasi-adiabatic continuation gives tighter bounds for the higher dimensional Lieb-Schultz-Mattis theorem.  On the other hand, the ``exact" quasi-adiabatic continuation is more convenient for the proof of Hall conductance quantization.  The exact form has the particular advantage that one may choose it so that evolution under the quasi-adiabatic continuation operator also obeys a Lieb-Robinson bound.

We omit all the details of error estimates in this review.

\subsection{Sketched Proof of Higher Dimensional Lieb-Schultz-Mattis Theorem}
We now sketch the proof of \cref{hlsm}.
The proof is variational, like the one-dimensional proof.  It is also by contradiction.  Let us assume that there is a gap $\Delta E$, and for large enough $\Delta E$ we will derive a contradiction.

Let $\psi_0$ be the ground state of $H$.  Since $H$ is translation invariant, $T\psi_0=z\psi_0$ for some $z$ with $|z|=1$.

Let $U$ be an operator that implements a quasi-adiabatic continuation of $H_{\theta,0}$ from
$\theta=0$ to $2\pi$, with the quasi-adiabatic parameter chosen appropriately (we do not go into the details) to make the following estimates work.

We emphasize that we do not make any assumption that $H_{\theta,0}$ has a gap for $\theta\neq 0$.  Indeed, by the arguments above, we know that the gap closes at some $\theta$.
Nevertheless, we define $U$ by integrating the quasi-adiabatic evolution operator as if there were a gap $\Delta E$.

Our variational state will be $\Phi=U\psi_0$.  This is similar to the one-dimensional construction with the operator $U_{LSM}$ replaced with a quasi-adiabatic evolution.
As in the one-dimensional proof, we prove two things: we bound $\langle \Phi,H \Phi\rangle-\langle \psi_0,H \psi_0\rangle$, and we compute $\langle \Phi,T \Phi\rangle$ to show that $\Phi$ is orthogonal to $\psi_0$.

To bound $\langle \Phi,H \Phi\rangle-\langle \psi_0,H \psi_0\rangle$, write
$$H=H_1+H_2,$$
where $H_1$ is the sum of terms $h_X$ such that $X$ is closer to the set of sites with first coordinate $0$ than it is to the set of sites with first coordinate $L/2$, and $H_2$ is the sum of the remaining terms:
$$H_1\equiv \sum_{X \mathrm{closer\, to\,} 0} h_X$$
and $$H_2 \equiv \sum_{\mathrm{remaining\,} X} h_X.$$

The term $H_2$ commutes with $U$ up to exponentially small (in $L\Delta E/v_{LR}$) error by locality of the quasi-adiabatic evolution operator.  At the same time, $\Vert H_2 \Vert$ is only polynomially large in $L$, and so for
 $\Delta E$ sufficiently large compared to $\log(L)/L$,
the commutator of the second term with $U$ is polynomially small (and indeed smaller than $\log(L)/L)$.

To estimate $\langle \Phi,H_1 \Phi\rangle-\langle \psi_0,H_1 \psi_0\rangle$, define $U'$ to implement quasi-adiabatic evolution of $H_{0,\theta'}$ as $\theta'$ goes from $0$ to $-2\pi$.  
Define $W$ to implement quasi-adiabatic evolution of $H_{\theta,-\theta}$ as $\theta$ goes from $0$ to $2\pi$.  Note the signs!

Since this is a sketched proof, we will write $\approx$ to indicate that something holds up to a polynomial in $L$ times something exponentially small in $L\Delta E/v_{LR}$, so that for $\Delta E$ sufficiently large compared to $\log(L)/L$, this $\approx$ indicates a polynomially small error.
One may show the following:
\begin{align}
\label{eseq}
\langle \Phi,H_1 \Phi\rangle  \approx &\langle U' \Phi,H_1 U' \Phi\rangle \\ \nonumber
\approx & \langle W \psi_0, H_1 W\psi_0 \rangle \\ \nonumber
\approx & \langle \psi_0, H_1 \psi_0\rangle,
\end{align}
where the first line is by the same locality of quasi-adiabatic evolution argument as we used in considering the commutator $[H_2,U]$.
The second line of \cref{eseq} is from $U' U \approx W$: this result can be understood as the quasi-adiabatic evolution operator that generates $W$ is a sum of two operators, one coming from the change in
$H_{\theta,\theta'}$ with respect to $\theta$ and the other the change with respect to $\theta'$, while the operators that generate $U,U'$ respectively come from the change
in $H_{\theta,\theta'}$ with respect to either $\theta$ or $\theta'$.  This simple argument doesn't fully justify the approximate equality of course, as the evolutions are taken simultaneously in $W$ ($\theta$ goes to $2\pi$ while $\theta'$ goes to $-2\pi$ in $W$) and sequentially in $U' U$, but using locality one may show it is approximately true.
The third line follows from \cref{gaugeinvar}: since $H_{\theta,-\theta}$ is unitarily equivalent to $H$, the Hamiltonian $H_{\theta,-\theta}$ must also have the same gap $\Delta E$ and so quasi-adiabatic evolution will approximately evolve the ground state of $H_{0,0}$ to the ground state $H_{2\pi,-2\pi}=H_{0,0}$, up to some phase.  
That is, while the gap will close for $H_{\theta,0}$, it remains open for $H_{\theta,-\theta'}$.
At this point in the proof, the phase is not important, since it cancels, but in the next step a similar phase will be important.  

This completes the sketch of the proof that $\langle \Phi,H\Phi\rangle-\langle \psi_0,H\psi_0\rangle$ is small.  We now sketch the proof that $\Phi$ has small overlap with $\psi_0$.  The ground state $\psi_0$ is an eigenvector of $T$ with some eigenvalue $z$.
We consider
$z^{-1}\langle \Phi,T \Phi\rangle$ and bound it away from $1$.  We have
\begin{align}
z^{-1}\langle \Phi,T \Phi\rangle =&\langle U \psi_0, T U T^{-1} \psi_0\rangle \\ \nonumber
\approx & \langle U' U \psi_0, U' \Bigl( T U T^{-1}\Bigr) \psi_0\rangle.
\end{align}
We have, as discussed above, $U' U \psi_0 \approx W \psi_0$, which is approximately equal to $\psi_0$ up to some phase.  A similar result holds for  $U' \Bigl( T U T^{-1}\Bigr)$:
this operator can be considered as describing quasi-adiabatic evolution in a family of Hamiltonians where instead the parameter $\theta$ describes a gauge field near the line with first coordinate $1$, rather than first coordinate $0$, while the parameter $\theta'$ still describes a gauge field near the line with first coordinate $L/2$.
So, $U' \Bigl( T U T^{-1}\Bigr)\psi_0$ is also equal to $\psi_0$ up to a phase.  However, analyzing this phase (which is approximately the geometric phase of some adiabatic evolution) shows that the two phases differ if $\langle \psi_0,Q \psi_0\rangle/L$ is non-integer, giving the desired result.

Remark: note that if $\langle \psi_0,Q \psi_0\rangle/L$ is non-integer, then its difference from the nearest integer is $\Omega(1/L)$ since $Q$ has integer eigenvalues.  So, even if the difference from $\langle \psi_0,Q \psi_0\rangle/L$ to the nearest integer is $o(1)$, one may still bound the errors terms to show that the two phases differ.

\section{Hall Conductance Quantization}
\label{qhe}
\subsection{Introduction}
In 1879, Edwin Hall performed an experiment in an attempt to determine the sign of the charge of charge carriers in a metal.  Was current caused by negative charge carriers flowing in one direction or by positive charge carriers flowing in the opposite direction?  Consider a sample of some metal, which looks like a rectangle as viewed from above.
He ran a current from the left side of the rectangle to the right side, while applying a magnetic field into the plane.  Maxwell's equations predicted that the charge carriers would experience a force, determined by their electric charge times the cross product of their velocity with the magnetic field.  This force is in the plane of the rectangle, and perpendicular to the direction of current.  The sign of this force is unchanged if one changes both the sign of the charge carriers and the sign of their velocity.  This force is expected to lead to an accumulation of the charge carriers at the top or bottom of edge of the rectangle, leading to a voltage between the top and bottom edge.  This effect, that a magnetic field can lead to a voltage perpendicular to the current, is called the Hall effect.

The sign of the voltage for most materials agrees with what one would expect for charge carriers with a negative electric charge (i.e., electrons), though in some semiconductors, the sign is reversed and it is more natural to think of holes in the band as carrying the charge.

The Hall conductance has units of 
$$\frac{\mathrm{current}}{\mathrm{voltage}}=\frac{{\mathrm{charge}}}{{\mathrm {time}}} \cdot \frac{\mathrm{charge}}{\mathrm{energy}}=\frac{\mathrm{charge}^2}{\mathrm{time}\cdot\mathrm{energy}},$$
so that it has the same units as $e^2/h$ where $e$ is the charge of the electron and $h$ is Planck's constant.

Surprisingly, in 1980, von Klitzing found experimentally that, in two-dimensional semiconductors at low temperatures and large magnetic field, the Hall conductance was quantized in integer multiples of $e^2/h$ to very high accuracy\footnote{There is also a fractional quantum Hall effect, where the Hall conductance is a rational multiple of $e^2/h$.  This can occur if the ground state is (approximately) degenerate.}.
The fundamental physical argument for this quantization was given by Laughlin\cite{laughlin1981quantized} but a mathematical proof remained open.

One surprising feature is that this very accurate quantization persists even though the actual physical samples were disordered.
In \cite{bellissard1994noncommutative}, noncommutative geometry techniques
were used to prove Hall conductance quantization for free (i.e., noninteracting) electrons with disorder.

\begin{figure}
\includegraphics[width=2in]{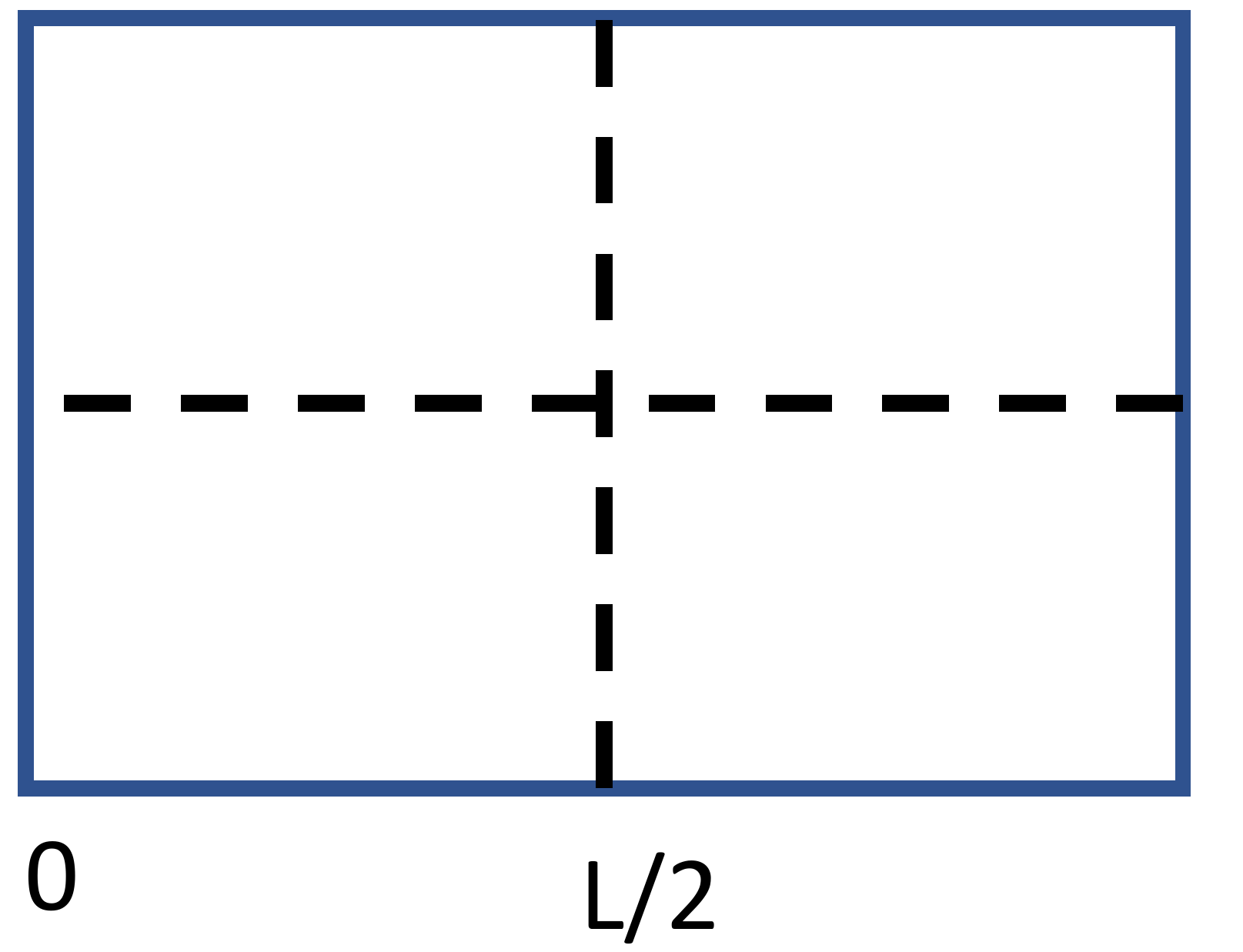}
\caption{Schematic illustration of two-dimensional system.  Individual sites are not shown.  Left and right sides of the figure are identified to give a cylinder, as well as top and bottom sides.  Numbers indicate first coordinate.  Twists $\theta,\phi$ twist terms near the left side and top side of the figure, respectively and are used in the averaging proof of \cite{avron1985quantization}.  Twists $\theta',\phi'$ twist terms near the dashed lines.  Twists $\theta',\phi'$ are used as a technical trick in the proof of \cite{hastings2015quantization}.}
\label{figqhe}
\end{figure}

In \cite{avron1985quantization}, Avron and Seiler
made another important advance, proving Hall conductance quantization under a certain averaging assumption.  They considered a system on a torus
and introduced two fluxes, $\theta,\phi$ on a longitude and meridian of the torus respectively.  See \cref{figqhe}.
We write $H_{\theta,\phi}$ to denote a Hamiltonian as a function of these two parameters; both $\theta$ and $\phi$ are periodic modulo $2\pi$.
The space of parameters $\theta,\phi$ is called the flux torus.

They assume that $H_{\theta,\phi}$ has a unique ground state for all $\theta,\phi$.  Consider adiabatically transporting the ground state around some infinitesimal loop in the flux torus near some given $\theta,\phi$.  The ground state acquires some Berry phase.  This Berry phase is related, by the Kubo formula, to the quantum Hall conductance at that $\theta,\phi$.
Physically, we can understand this relation as follows.  If we imagine changing one parameter (say, $\theta$), this corresponds to a changing magnetic field which induces a voltage.  This voltage induces a perpendicular current proportional to the Hall conductance, and this current will couple to the other parameter (in this case, $\phi$), changing the phase of the wavefunction.

The problem then is reduced to computing the Berry phase around such an infinitesimal loop, i.e., computing a Berry curvature.  The average of this curvature over the torus is quantized in integer multiples of $(2\pi)^{-1}$.  Thus, Avron and Seiler proved that the average of the Hall conductance over the torus is quantized.

However, this left open the question of quantization of the Hall conductance at a specific value of $\theta,\phi$, assuming only a spectral gap at that $\theta,\phi$.

\subsection{Results}
We now state the results of \cite{hastings2015quantization}, which proved Hall conductance quantization for interacting electrons without the averaging assumption.
The results are quantitative, giving error bounds that decay almost exponentially fast as $L \rightarrow \infty$.
A function $f$ is called \emph{almost-exponentially decaying}, if for all $c$ with $0 \leq c <1$ there is a constant $C$ such that $f(x) \leq C \exp(-x^c)$ for all sufficiently large $x$, and a quantity is \emph{almost-exponentially small} if it is bounded by an almost-exponentially decaying function. 

We consider a two-dimensional quantum system, with
sites on a torus $T$, with sites labeled by a pair $(i,j)$ periodic modulo $L$ for some $L$.  
We assume that there is a conserved charge $Q=\sum_v q_v$ as in the Lieb-Schultz-Mattis theorem and that the Hamiltonian has bounded strength and range.
We assume that the Hamiltonian has a unique groundstate, with a spectral gap at least $\Delta E$.

Specifically, it is proven:
\begin{theorem}
For any fixed, $L$-independent $R,J, \qmax$ and spectral gap $\Delta E>0$, for any Hamiltonian satisfying the above assumptions, the difference between the Hall conductance $\sigma_{xy}$ and the nearest integer multiple of $e^2/h$ is almost-exponentially small in $L$, where
$e^2/h$ denotes the square of the electron charge divided by Planck's constant.
\end{theorem}

The proof of this theorem takes several steps.
First, one replaces the Berry connection used to compute the Berry phase with the quasi-adiabatic evolution operator,
to relate the Hall conductance to the phase for quasi-adiabatic evolution around a small loop.
  One considers quasi-adiabatic evolution around a small, but not infinitesimal loop, on the flux torus near $(\theta,\phi)=(0,0)$.  Keeping the loop sufficiently small (the size of the loop polynomially small in the system size), the gap remains open on evolution on this loop.  Indeed, we may assume that the gap remains at least $(1-o(1)) \Delta E$, so that we may choose the function in
  quasi-adiabatic evolution so that it matches adiabatic evolution on this loop.  Thus, the ground state returns to itself after quasi-adiabatic evolution around this loop, up to a Berry phase.  For small loop size, this phase is proportional to the area times the Berry curvature plus higher order corrections in loop size.

Second, one considers various other choices of $(\theta,\phi)\neq (0,0)$.  For each such choice, one defines a path from $(0,0)$ to the given $(\theta,\phi)$,  around a small loop, and back to $(0,0)$.  Note that since we move now a large distance away from $(0,0)$, the gap may become small, or even vanish.  So, we do not have a guarantee that we return to the ground state at the end of the path.  However, we can make energy estimates to show that we indeed return to the ground state.  To do this, we use a similar trick to that done in the proof of the higher dimensional Lieb-Schultz-Mattis theorem.  In that proof, we introduced an additional twist $\theta'$ and used it to show that our variational state had energy close to the ground state.  Here we  introduce two extra twists $\theta',\phi'$, and we use them to show that the state at the end of the path has energy close to the ground state.  In this case, since we assume that the ground state is unique, this proves that we do return to the ground state up to small error.  Further, we may show that the phase acquired is approximately independent of the choice of $(\theta,\phi)$.

Next, one takes a product of the evolution over several such paths (each path going from $(0,0)$ to some $(\theta,\phi)\neq (0,0)$, around a loop, and back to $(0,0)$).  We do this such that the result is equivalent to evolution around a single large loop, i.e., certain segments of the paths cancel, leaving just the evolution around the large loop.
This large loop starts at $(0,0)$, then increases $\theta$ from $0$ to $2\pi$, keeping $\phi$ fixed.  Then it increases $\phi$ to $2\pi$, keeping $\theta$ fixed.  Then, it decreases $\theta$ from $2\pi$ to $0$, keeping $\phi$ fixed.  Finally, it decreases $\phi$ from $2\pi$ to $0$, keeping $\theta$ fixed.
Each of those four segments of the evolution approximately returns the ground state to itself, up to some phase; this again is shown by an energy argument.
However, using the $2\pi$ periodicity in the parameters $\theta,\phi$, the phases cancel.
Thus, the combined evolution gives a phase which is approximately an integer multiple of $2\pi$, and since this phase is approximately the product of the phases around the small loops, the phase for each small loop
is approximately an integer multiple of $(2\pi)^{-1}$ times the area of the loop.  This part of the proof is of course very similar to one way to show that the average of the Berry curvature over flux torus is quantized; essentially, it is a form of Stokes' theorem.  However, since we have used quasi-adiabatic evolution so that all the small loops contribute approximately the same phase, and since we have related the phase for the small loop near $(\theta,\phi)=(0,0)$ to the Berry curvature, it proves quantization without averaging assumption and without any assumption of the gap remaining open.

\bibliographystyle{emss}
\bibliography{refs.bib}

\appendix
\section{Lieb-Robinson Bounds}
\label{LRapp}
\subsection{Lieb-Robinson Bound}
Here we show
\begin{lemma}
Given operators $A$ supported on $X$ and $B$ supported on $Y$ with
$X\cap Y = \emptyset$ we have
\begin{align}
\label{CBseries}
&\Vert [A(t),B] \Vert \\ \nonumber\le&
2\Vert A \Vert \cdot \Vert B\Vert(2|t|)\sum_{Z_1:Z_1\cap X\ne\emptyset,Z_1\cap Y\ne\emptyset}
\Vert h_{Z_1}\Vert\\ \nonumber
&+2\Vert A \Vert\cdot \Vert B\Vert\frac{(2|t|)^2}{2!}
\sum_{Z_1:Z_1\cap X\ne\emptyset}\Vert h_{Z_1}\Vert
\sum_{Z_2:Z_2\cap Z_1\ne\emptyset,Z_2\cap Y\ne\emptyset}\Vert h_{Z_2}\Vert\\
\nonumber
&+2\Vert A \Vert \cdot\Vert B\Vert\frac{(2|t|)^3}{3!}
\sum_{Z_1:Z_1\cap X\ne\emptyset}\Vert h_{Z_1}\Vert
\sum_{Z_2:Z_2\cap Z_1\ne\emptyset}\Vert h_{Z_2}\Vert
\sum_{Z_3:Z_3\cap Z_2\ne\emptyset,Z_3\cap Y\ne\emptyset}\Vert h_{Z_3}\Vert\\ \nonumber &+\cdots
\end{align}
Remark: the $k$-th term of the above series is equal to
$2\Vert A \Vert \cdot\Vert B\Vert\frac{(2|t|)^k}{k!}$ times the sum over
sets $Z_1,\ldots,Z_k$ with $Z_1\cap X \ne \emptyset$, $Z_j\cap Z_{j+1}  \neq \emptyset$ for $0\le j <k$, and $Z_k\cap Y \neq \emptyset$, of the product
$\prod_{j=1}^k\Vert h_{Z_j}\Vert$.
\begin{proof}
We assume $t>0$ because negative $t$ can be treated in the same way. 
Let $\epsilon=t/N$ with a large positive integer $N$, and let 
$$t_n=\frac{t}{N}n\quad\mbox{for}\ n=0,1,\ldots, N.$$ 
Then we have 
\begin{equation}
\left\Vert[A(t),B]\right\Vert-\left\Vert[A(0),B]\right\Vert 
=\sum_{i=0}^{N-1}\epsilon\times
\frac{\left\Vert[A(t_{n+1}),B]\right\Vert-\left\Vert[A(t_n),B]\right\Vert}{\epsilon}.
\label{sumid} 
\end{equation}
In order to obtain the bound (\ref{Commnormbound}) below, 
we want to estimate the summand in the right-hand side. 
To begin with, we note that the identity,  
$\left\Vert U^\ast OU\right\Vert=\Vert O\Vert$, holds for any observable $O$ 
and for any unitary operator $U$. Using this fact, we have  
\begin{align}
\label{difnorm}
&\left\Vert[A(t_{n+1}),B]\right\Vert-\left\Vert[A(t_n),B]\right\Vert
\\ \nonumber
=&\left\Vert[A(\epsilon),B(-t_n)]\right\Vert-\left\Vert[A,B(-t_n)]\right\Vert\\ \nonumber
\le&\left\Vert[A+i\epsilon[h_\Lambda,A],B(-t_n)]\right\Vert
-\left\Vert[A,B(-t_n)]\right\Vert+{\mathcal O}(\epsilon^2)\\ \nonumber
=&\left\Vert[A+i\epsilon[I_X,A],B(-t_n)]\right\Vert
-\left\Vert[A,B(-t_n)]\right\Vert+{\mathcal O}(\epsilon^2)
\end{align}
with 
\begin{equation}
I_X=\sum_{Z:Z\cap X\ne \emptyset}h_Z,
\label{defIX}
\end{equation}
where we have used 
\begin{equation}
A(\epsilon)=A+i\epsilon[h_\Lambda,A]+{\mathcal O}(\epsilon^2)
\end{equation}
and a triangle inequality. Further, by using 
\begin{equation}
A+i\epsilon[I_X,A]=e^{i\epsilon I_X}Ae^{-i\epsilon I_X}+{\mathcal O}(\epsilon^2), 
\end{equation}
we have 
\begin{align}
\left\Vert[A+i\epsilon[I_X,A],B(-t_n)]\right\Vert
\le&\left\Vert[e^{i\epsilon I_X}Ae^{-i\epsilon I_X},B(-t_n)]\right\Vert
+{\mathcal O}(\epsilon^2)\\\nonumber
=&\left\Vert[A,e^{-i\epsilon I_X}B(-t_n)e^{i\epsilon I_X}]\right\Vert
+{\mathcal O}(\epsilon^2)\\\nonumber
\le&\left\Vert[A,B(-t_i)-i\epsilon[I_X,B(-t_n)]]\right\Vert
+{\mathcal O}(\epsilon^2)\\\nonumber
\le&\left\Vert[A,B(-t_n)]\right\Vert+\epsilon\left\Vert[A,[I_X,B(-t_n)]]\right\Vert
+{\mathcal O}(\epsilon^2). 
\end{align}
Substituting this into the right-hand side in the last line of (\ref{difnorm}), 
we obtain 
\begin{align}
\left\Vert[A(t_{n+1}),B]\right\Vert-\left\Vert[A(t_n),B]\right\Vert
\le&\epsilon\left\Vert[A,[I_X,B(-t_n)]]\right\Vert
+{\mathcal O}(\epsilon^2)\\ \nonumber
\le&2\epsilon\Vert A\Vert\left\Vert[I_X(t_n),B]\right\Vert
+{\mathcal O}(\epsilon^2).
\end{align}
Further substituting this into the right-hand side of (\ref{sumid}) and  
using (\ref{defIX}), we have 
\begin{align}
\left\Vert[A(t),B]\right\Vert-\left\Vert[A(0),B]\right\Vert 
\le&2\Vert A\Vert\sum_{n=0}^{N-1}\epsilon\times\left\Vert[I_X(t_n),B]\right\Vert
+{\mathcal O}(\epsilon)\\ \nonumber
\le&2\Vert A\Vert\sum_{Z:Z\cap X\ne \emptyset}\sum_{n=0}^{N-1}\epsilon\times
\left\Vert[h_Z(t_n),B]\right\Vert+{\mathcal O}(\epsilon). 
\end{align}
Since $h_Z(t)$ is a continuous function of the time $t$ for a finite volume, 
the sum in the right-hand side converges to the integral in 
the limit $\epsilon\downarrow 0$ (i.e., $N\uparrow\infty$) 
for any fixed finite lattice $\Lambda$. In consequence, we obtain
\begin{equation}
\left\Vert[A(t),B]\right\Vert-\left\Vert[A(0),B]\right\Vert 
\le 2\Vert A\Vert\sum_{Z:Z\cap X\ne \emptyset}\int_0^{|t|}ds 
\left\Vert[h_Z(s),B]\right\Vert.
\label{Commnormbound} 
\end{equation}

We define 
\begin{equation}
C_B(X,t):=\sup_{A\in{\mathcal A}_X}\frac{\Vert[A(t),B]\Vert}{\Vert A\Vert},
\label{CBXt}
\end{equation}
where ${\mathcal A}_X$ is the algebra of observables supported on
the set $X$. Then we have
\begin{equation}
C_B(X,t)\le C_B(X,0)+2\sum_{Z:Z\cap X\ne \emptyset}\Vert h_Z\Vert
\int_0^{|t|}ds\> C_B(Z,s)
\label{CBXtbound}
\end{equation}
from the above bound (\ref{Commnormbound}).
Assume $\dist(X,Y)>0$. 
Then we have $C_B(X,0)=0$ from the definition of $C_B(X,t)$, and note that 
\be
C_B(Z,0)\le 2 \Vert B \Vert,
\ee
for $Z\cap Y\ne\emptyset$ and
\be
C_B(Z,0))=0
\ee
otherwise.
Using these facts and the above bound (\ref{CBXtbound}) iteratively, 
we obtain 
\begin{align}
\nonumber
C_B(X,t)\le&2\sum_{Z_1:Z_1\cap X\ne\emptyset}\Vert h_{Z_1}\Vert
\int_0^{|t|}ds_1\> C_B(Z_1,s_1)\\ \nonumber
\le&2\sum_{Z_1:Z_1\cap X\ne\emptyset}\Vert h_{Z_1}\Vert
\int_0^{|t|}ds_1\> C_B(Z_1,0)\\ \nonumber
&+2^2\sum_{Z_1:Z_1\cap X\ne\emptyset}\Vert h_{Z_1}\Vert
\sum_{Z_2:Z_2\cap Z_1\ne\emptyset}\Vert h_{Z_2}\Vert
\int_0^{|t|}ds_1\int_0^{|s_1|}ds_2\> C_B(Z_2,s_2)\\ \nonumber
\\ \nonumber
 \leq &\cdots
\end{align}
So,
\begin{align}
&
C_B(X,t)\\ \nonumber\le&
2\Vert B\Vert(2|t|)\sum_{Z_1:Z_1\cap X\ne\emptyset,Z_1\cap Y\ne\emptyset}
\Vert h_{Z_1}\Vert\\ \nonumber
&+2\Vert B\Vert\frac{(2|t|)^2}{2!}
\sum_{Z_1:Z_1\cap X\ne\emptyset}\Vert h_{Z_1}\Vert
\sum_{Z_2:Z_2\cap Z_1\ne\emptyset,Z_2\cap Y\ne\emptyset}\Vert h_{Z_2}\Vert\\
\nonumber
&+2\Vert B\Vert\frac{(2|t|)^3}{3!}
\sum_{Z_1:Z_1\cap X\ne\emptyset}\Vert h_{Z_1}\Vert
\sum_{Z_2:Z_2\cap Z_1\ne\emptyset}\Vert h_{Z_2}\Vert
\sum_{Z_3:Z_3\cap Z_2\ne\emptyset,Z_3\cap Y\ne\emptyset}\Vert h_{Z_3}\Vert+\cdots
\end{align}
\end{proof}
\end{lemma}

\end{document}